\documentclass[reprint, aps, prl, twocolumn, 
superscriptaddress
, showpacs, preprintnumbers, amsmath, amssymb, floatfix]{revtex4-1}

\usepackage{import}
\usepackage{graphicx}
\usepackage[mathlines]{lineno}
\usepackage{xspace} 
\usepackage{lipsum}
\usepackage{float}
\usepackage[caption = false]{subfig}
\usepackage{placeins}
\usepackage{hyperref}

\begin{document}



\title{Imaging the Oblique Propagation of Electrons in Germanium Crystals\\
 at Low Temperature and Low Electric Field}

\author{R. A. Moffatt}
\email{rmoffatt@stanford.edu}
\affiliation{Dept. of Physics, Stanford University, Stanford, CA 94305, USA}

\author{B. Cabrera}
\affiliation{Dept. of Physics, Stanford University, Stanford, CA 94305, USA}

\author{B. M. Corcoran}
\affiliation{Dept. of Physics, Stanford University, Stanford, CA 94305, USA}

\author{J. M. Kreikebaum}
\affiliation{Dept. of Physics, Stanford University, Stanford, CA 94305, USA}

\author{P. Redl}
\affiliation{Dept. of Physics, Stanford University, Stanford, CA 94305, USA}

\author{B. Shank}
\affiliation{Dept. of Physics, Stanford University, Stanford, CA 94305, USA}

\author{J. J. Yen}
\affiliation{Dept. of Physics, Stanford University, Stanford, CA 94305, USA}

\author{B. A. Young}
\affiliation{Dept. of Physics, Stanford University, Stanford, CA 94305, USA}
\affiliation{Dept. of Physics, Santa Clara University, Santa Clara, CA 95053, USA}

\author{P. L. Brink}
\affiliation{SLAC National Accelerator Facility, Menlo Park, CA 94025, USA}

\author{M. Cherry}
\affiliation{SLAC National Accelerator Facility, Menlo Park, CA 94025, USA}

\author{A. Tomada}
\affiliation{SLAC National Accelerator Facility, Menlo Park, CA 94025, USA}


\author{A. Phipps}
\affiliation{Department of Physics, University of California, Berkeley, CA 94720, USA}

\author{B. Sadoulet}
\affiliation{Department of Physics, University of California, Berkeley, CA 94720, USA}

\author{K. M. Sundqvist}
\thanks{Present Address: Electrical and Computer Engineering, Texas A\&M University, College Station, TX 77843, USA}
\affiliation{Department of Physics, University of California, Berkeley, CA 94720, USA}

\date{\today}

\begin{abstract}


Excited electrons in the conduction band of germanium collect into four energy minima, or valleys, in momentum space. These local minima have highly anisotropic mass tensors which cause the electrons to travel in directions which are oblique to an applied electric field at sub-Kelvin temperatures and low electric fields, in contrast to the more isotropic behavior of the holes. This experiment produces, for the first time, a full two-dimensional image of the oblique electron and hole propagation and the quantum transitions of electrons between valleys for electric fields oriented along the [0,0,1] direction. Charge carriers are excited with a focused laser pulse on one face of a germanium crystal and then drifted through the crystal by a uniform electric field of strength between 0.5 and 6 V/cm. The pattern of charge density arriving on the opposite face is used to reconstruct the trajectories of the carriers. Measurements of the two-dimensional pattern of charge density are compared in detail with Monte Carlo simulations developed for the Cryogenic Dark Matter Search (SuperCDMS) to model the transport of charge carriers in high-purity germanium detectors.

\end{abstract}

\pacs{72.10.Di, 72.20.Jv, 72.20.-i, 72.20.Dp, 72.80.Cw, 71.18.+y}

\maketitle


\section{Introduction}


Because germanium is an indirect-gap semiconductor, the energy minimum of its conduction band does not occur at zero momentum. Rather, the conduction band has four minima, or valleys, located at the edges of the Brillouin zone in the four $\left<1,1,1\right>$ crystal directions. The effective masses of the electrons in these valleys are highly anisotropic, and take on values of 1.58 and 0.081 times the mass of the electron in vacuum in the directions parallel and perpendicular to the $\left<1,1,1\right>$ directions, respectively. \cite{Jacoboni1983,Cabrera2010} Because of the anisotropic mass, each electron's acceleration is rarely parallel to the applied force; the direction of acceleration depends on the mass tensor of the valley the electron is occupying. As a result, an initially localized group of electrons in a uniform electric field will spatially separate into four clusters, each consisting of electrons occupying one of the four valleys.

At high temperatures, or in the presence of high electric fields or impurity concentrations, electrons will undergo frequent quantum transitions between these four valleys, resulting in a nearly isotropic electron mobility. However, if the temperature, electric field, and impurity concentrations are all sufficiently low, the mean-free-path for inter-valley scattering can exceed the sample size, and the mass anisotropy will become apparent. This effect was the cause of a surprising asymmetry observed between the collection of electrons and holes in the germanium crystals used as dark matter detectors in the Cryogenic Dark Matter Search (CDMS). \cite{Pyle2012} To model this asymmetry, we developed Monte Carlo simulations of charge transport in our germanium crystals. The following experiment was designed to verify these simulations in detail by directly imaging the propagation of charge carriers.






\begin{figure}[htb]
\begin{center}
\subfloat[Illustration of scanning system.]{\includegraphics[width=0.55\linewidth]{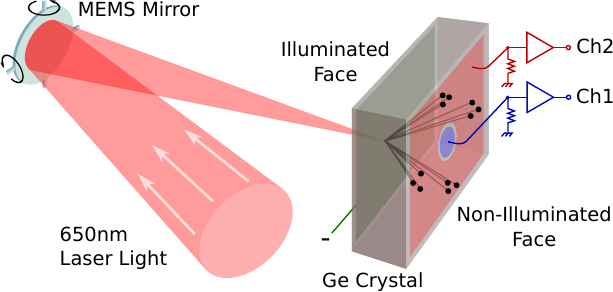}}\\
\subfloat[Photograph of a raster scan.]{\includegraphics[width=0.55\linewidth]{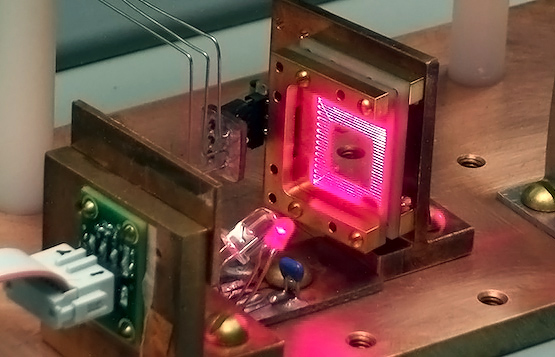}}
\caption{{\bf Laser Scanning System:} (a) Schematic of the crystal sample and laser scanning system. (b) Benchtop photograph of the scanning system with a raster-scan across the surface of the crystal and the copper-clad G10 crystal holder, shown here outside of the He-3 cryostat. Copper shielding removed for photograph.}
\label{fig:schematic}
\end{center}
\end{figure}


\vspace{-1cm}
\section{Experimental Setup}


The crystal under test was cut from a 3.89mm thick wafer of high purity germanium.\cite{CrystalProperties} The front and back faces are 1cm$\times$1cm, and lie in the $(0,0,1)$ crystal plane, while the sides lie in the $(1,1,0)$ and $(1,\overline{1},0)$ planes. The illuminated (front) face of the crystal is patterned with an aluminum-tungsten mesh electrode, with 10$\mu$m pitch and 20\% coverage.\cite{Trilayer} The non-illuminated (back) face is covered by two solid electrodes, one of which is circular with a diameter of 160$\mu$m, and separated by a 10$\mu$m gap from the other electrode, which covers the rest of the face of the crystal. Both electrodes on the non-illuminated face are connected to ground through 10.8M$\Omega$ resistors.

A benchtop photograph of the crystal sample is shown in Fig. \ref{fig:schematic}b. The sample is mounted to a copper baseplate (bottom of photograph), and is cooled to 600mK under vacuum in a He-3 cryostat. The crystal sample is shielded from both electrical noise and 4K thermal radiation by a copper enclosure mounted on the baseplate.

To generate free charge carriers, the illuminated face is exposed to a 20$\mu$W, 100ns pulse of 650nm laser light, focused to a 60$\mu$m diameter spot, which creates a cloud of electron-hole pairs near the illuminated surface. Carriers are propagated through the bulk of the crystal in the uniform electric field induced by the DC bias voltage of the mesh electrode on the illuminated face. The sign of the bias voltage determines whether electrons or holes propagate through the bulk of the crystal. Once collected by the electrodes on the non-illuminated face, the charge carriers produce proportional pulses in voltage, which are measured by high-impedance MESFET amplifiers, mounted on the 4K stage in the He-3 cryostat.\cite{Lee1989,Moffatt2014}

The two-dimensional pattern of charge density is determined by measuring the charge collected by the small, circular electrode as a function of the position of the excitation point on the illuminated face. Because translation of the excitation point causes a corresponding translation of the charge density pattern, this procedure is equivalent to keeping the excitation point fixed, while moving the small, circular electrode to probe different regions of the charge density pattern. The position of the excitation point is controlled by means of a Micro Electro-Mechanical System (MEMS) mirror from Mirrorcle Technologies, Inc.\cite{Mirrorcle} The MEMS mirror can be tilted along two axes by computer control, and was modified by the vendor for operation below 1K.


\section{Results}

\begin{figure}[htb]
\begin{center}
\subfloat[Hole Data]{\includegraphics[width=0.74\linewidth]{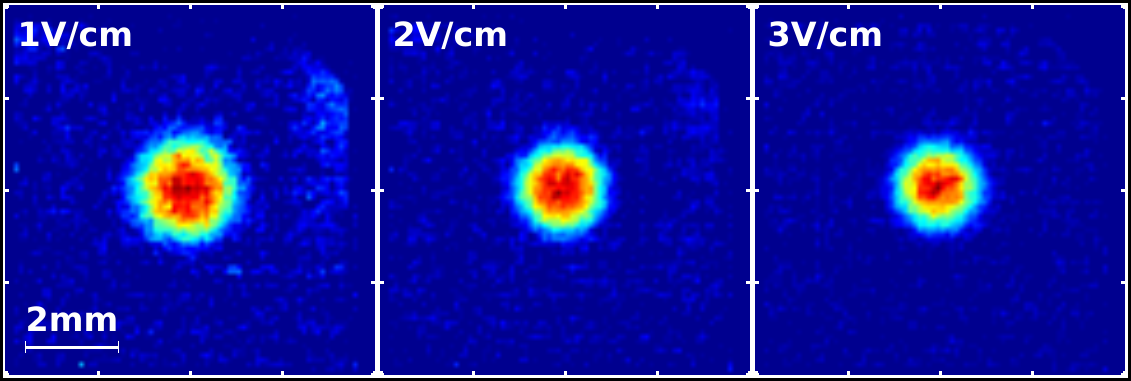}}\\
\subfloat[Hole Simulation]{\includegraphics[width=0.74\linewidth]{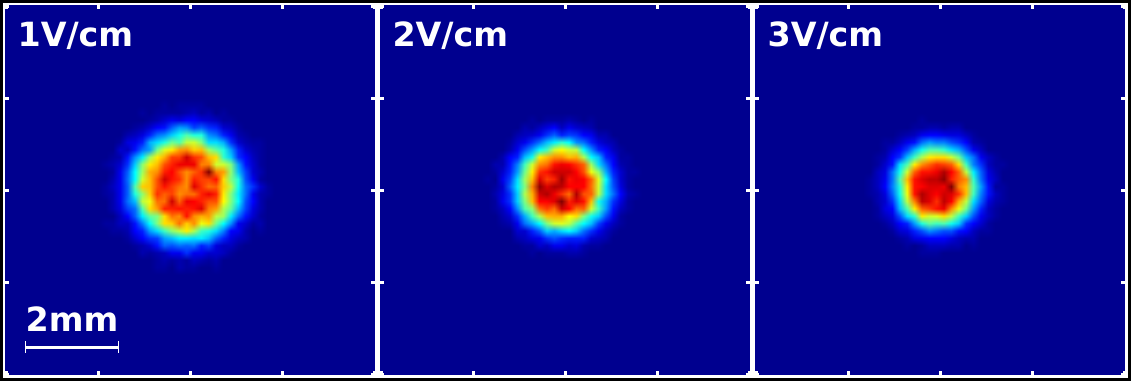}}\\
\subfloat[Data (solid) vs. Simulation (dotted)]{\includegraphics[width=0.74\linewidth]{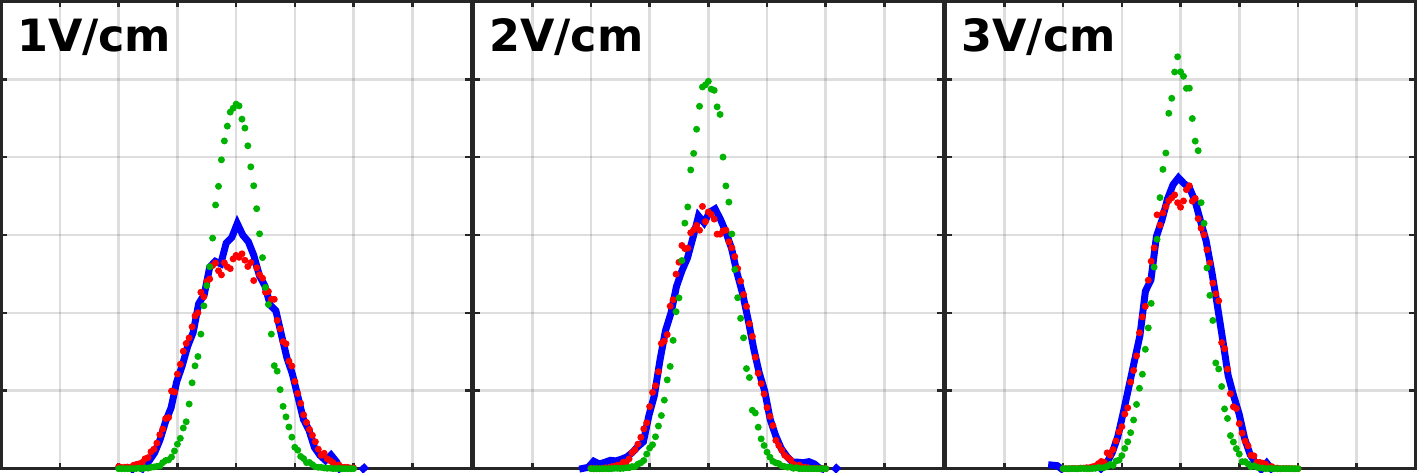}}
\caption{{\bf Hole Charge Density Patterns:} (a): Data. (b): Simulation. (c): One-dimensional projection of charge density onto x-axis. The data (solid, blue) are compared to the simulation incorporating electrostatic repulsion (dotted, red), and the simulation with no repulsion (dotted, green). The horizontal scale ranges from -4mm to +4mm. The vertical scale is arbitrary.}
\label{fig:Holes}
\end{center}
\end{figure}
\begin{figure}[htb]
\begin{center}
\subfloat[Electron Data]{\includegraphics[width=\linewidth]{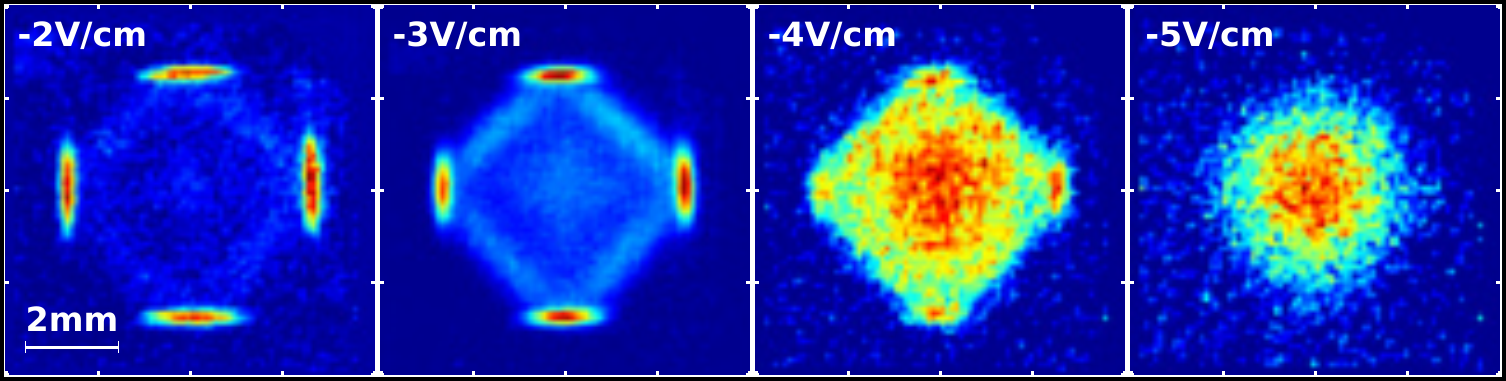}}\\ 
\subfloat[Electron Simulation (Redl)]{\includegraphics[width=\linewidth]{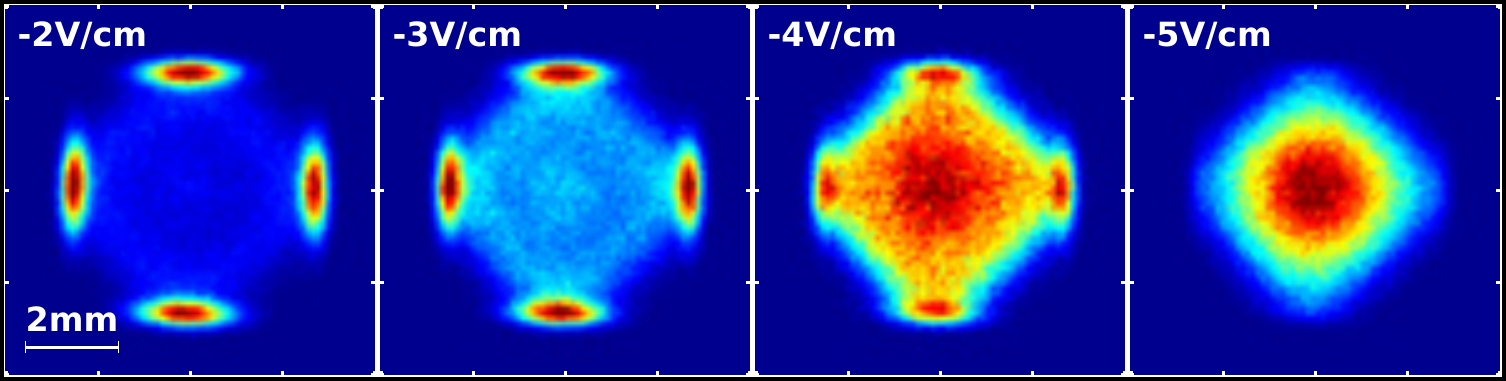}}\\
\subfloat[Data (solid blue) vs. Simulation (dotted red)]{\includegraphics[width=\linewidth]{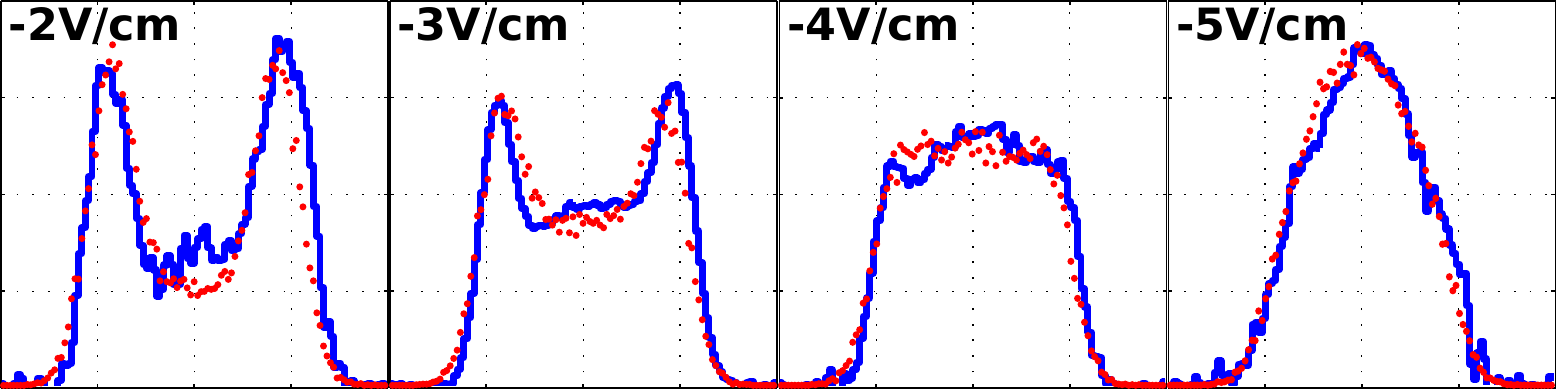}}
\caption{{\bf Electron Charge Density Patterns:} (a): Data. (b): Redl simulation. (c): One-dimensional projection of charge density onto a diagonal axis. The data (solid, blue) are compared to the Redl simulation employing the Herring-Vogt approximation (dotted, red). The horizontal scale ranges from -4mm to +4mm. The vertical scale is arbitrary.}
\label{fig:Electrons}
\end{center}
\end{figure}

The experimental results are shown in Figs. \ref{fig:Holes}a and \ref{fig:Electrons}a, which plot the voltage amplitude of the small circular electrode as a function of the position of the point of laser excitation. These data provide a map of the two-dimensional pattern of charge arriving on the non-illuminated face of the crystal, and are compared to the simulated distribution of charge carriers calculated by Monte Carlo simulation, shown in Figs. \ref{fig:Holes}b and \ref{fig:Electrons}b.


The comparison between the Monte Carlo simulation and the experimental data can be more clearly seen in Figs. \ref{fig:Holes}c and \ref{fig:Electrons}c, which show one-dimensional projections of the two-dimensional charge density patterns, for both the data (solid line) and the simulation (dotted line), in the presence of several different electric field strengths.

The effects of the electron mass anisotropy are clearly evident in Fig. \ref{fig:Electrons}a. At low electric fields (less than $\sim$3V/cm), the electrons separate into four groups, each corresponding to one of the conduction band minima, or valleys. At high electric fields (greater than $\sim$4V/cm), the inter-valley scattering induced by phonon emission washes out the effects of the anisotropic mass, and the charge density pattern approaches a Gaussian distribution.


\begin{figure}[htb]
\begin{center}
\includegraphics[width=\linewidth]{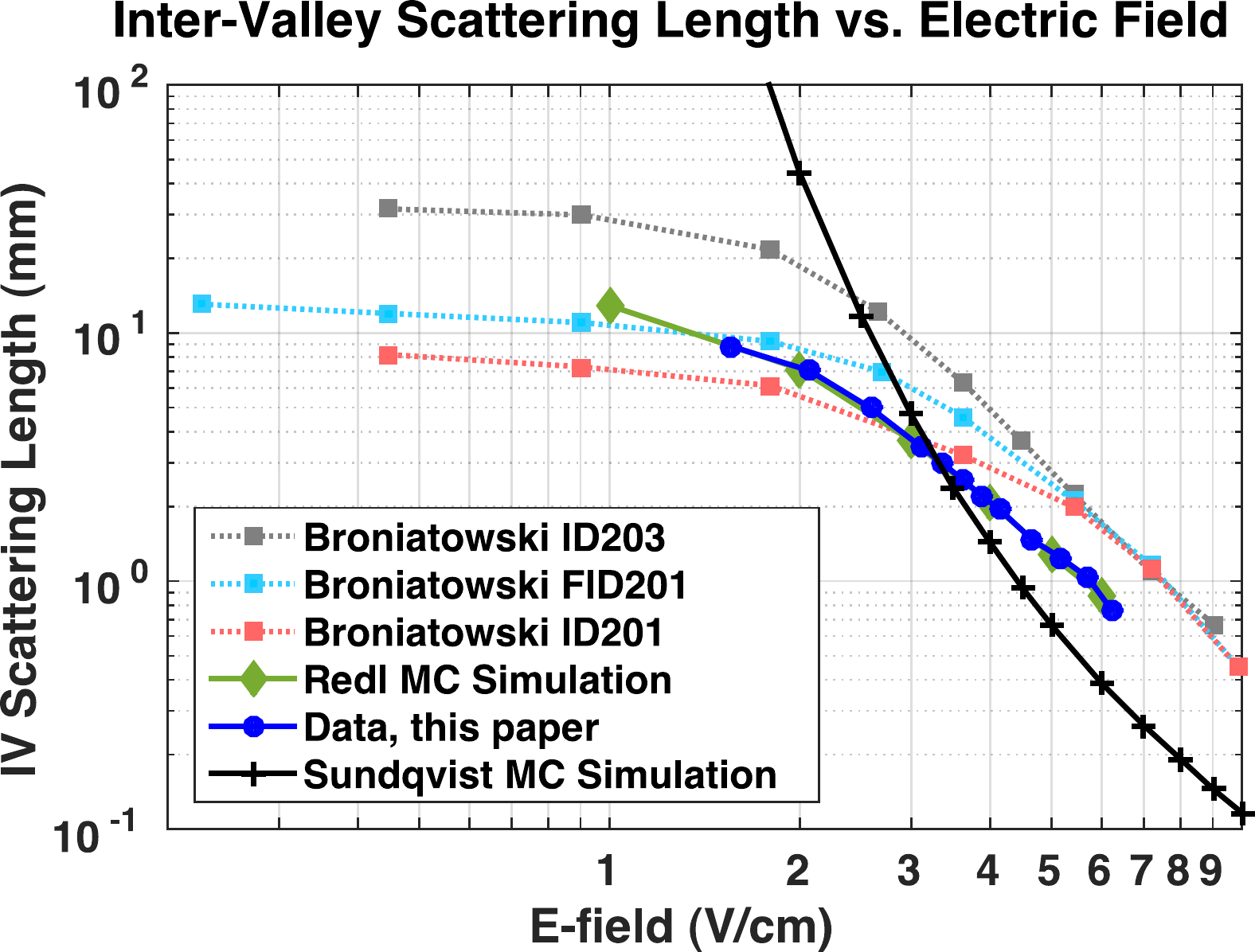}
\caption{{\bf Electron inter-valley scattering length}, projected along $\left[ 0, 0, 1 \right]$, as a function of electric field strength. The data (blue) are compared to the Redl simulation (green) and the Sundqvist simulation (black). Also included is the IV scattering length calculated using data from references \cite{Broniatowski2014} and \cite{Piro2014}.}
\vspace{-0.4cm}
\label{fig:Inter-valleyScattering}
\end{center}
\end{figure}

\begin{figure}[htb]
\begin{center}
\subfloat[Redl Simulation]{\includegraphics[width=0.32\linewidth]{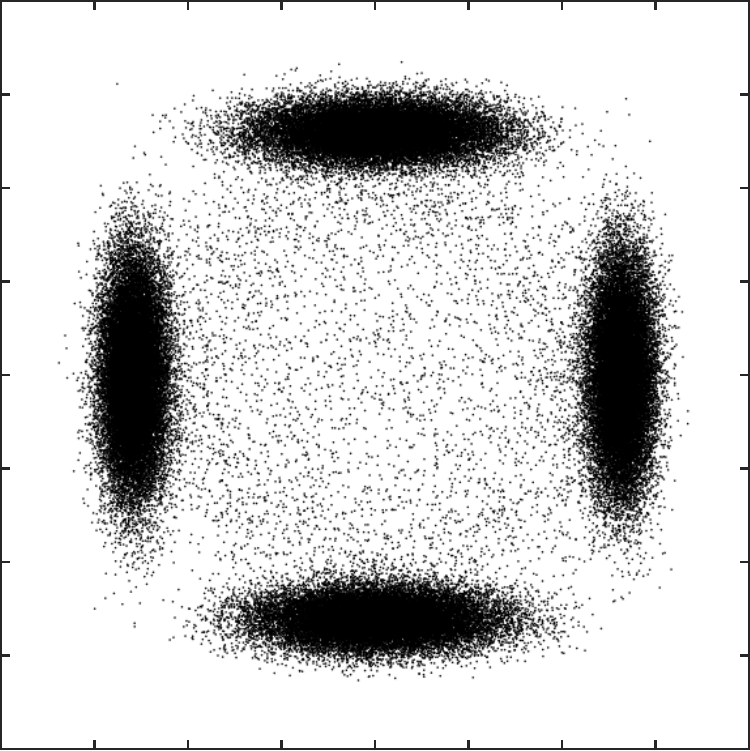}}\ 
\subfloat[Sundqvist Simulation]{\includegraphics[width=0.32\linewidth]{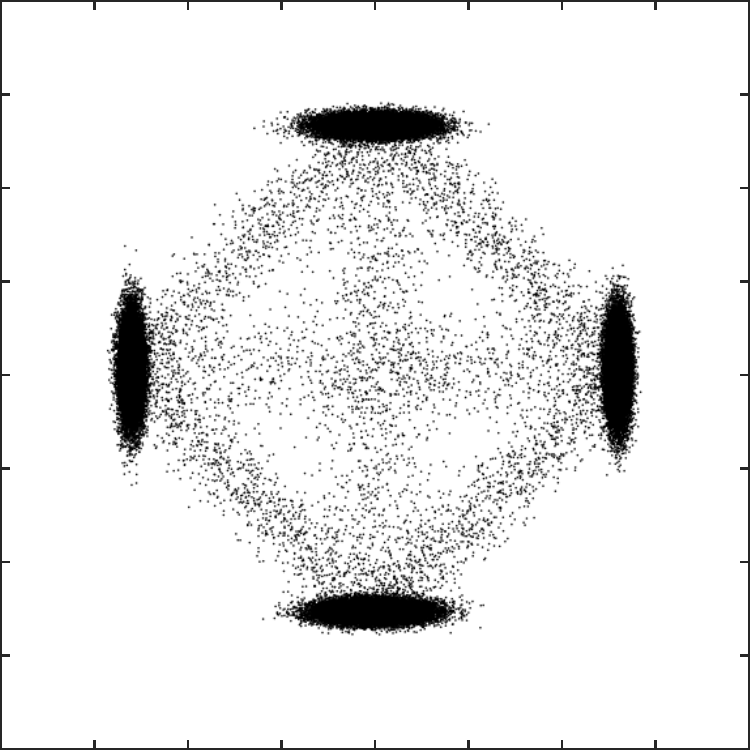}}\ 
\subfloat[Sundqvist Sim. with Repulsion]{\includegraphics[width=0.32\linewidth]{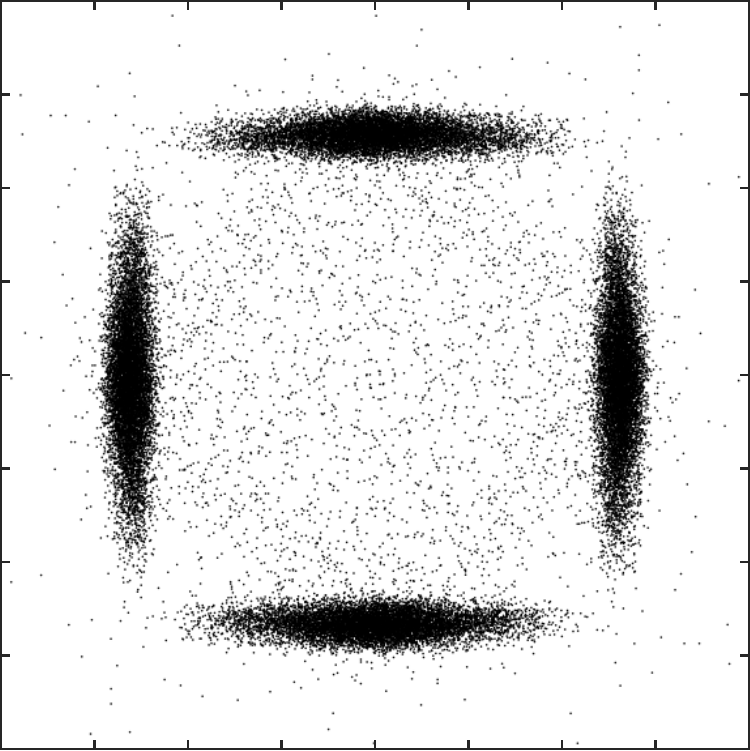}}
\caption{{\bf Comparison of Redl and Sundqvist Simulations:} (a), (b), and (c) show scatter plots of the final x and y electron locations in three simulations. The axes extend from -4mm to +4mm in the x and y-directions. (a) Reld simulation using Herring-Vogt approximation, which does not conserve momentum. (b) Sundqvist simulation, which correctly conserves momentum during phonon scattering.\cite{Sundqvist2012,Phipps2012,Sundqvist2014} (c) Sundqvist simulation with correction for electrostatic repulsion. For all simulations, the electric field strength was 2V/cm, and no impurity scattering was included.}
\vspace{-0.5cm}
\label{fig:SimulationComparison}
\end{center}
\end{figure}

\begin{figure}[htb]
\begin{center}
\subfloat{\includegraphics[width=0.7\linewidth]{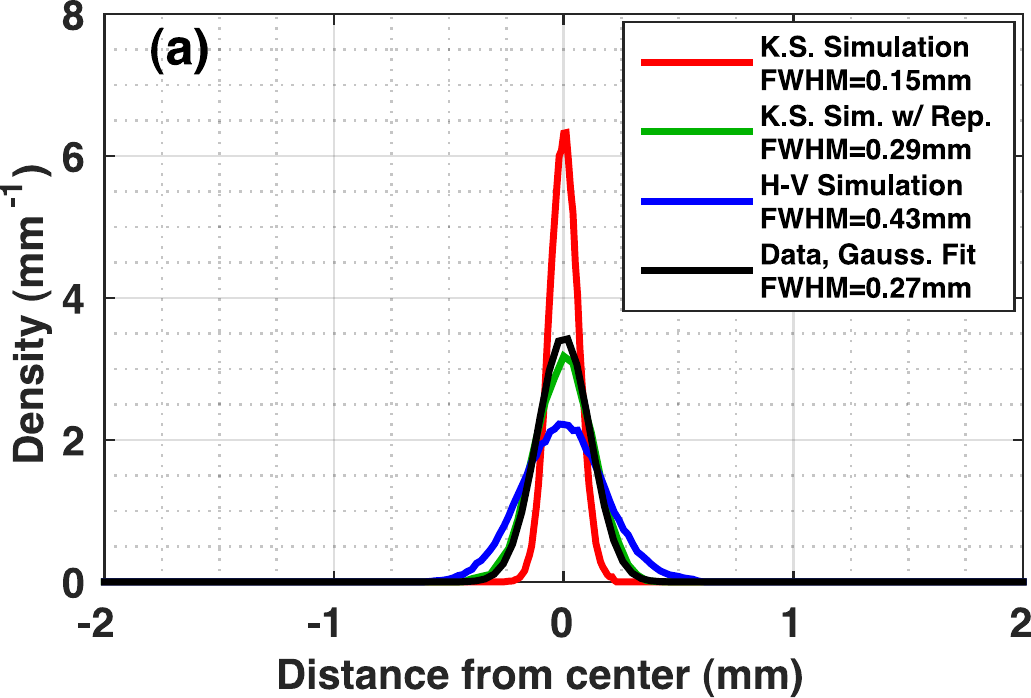}}\\
\subfloat{\includegraphics[width=0.7\linewidth]{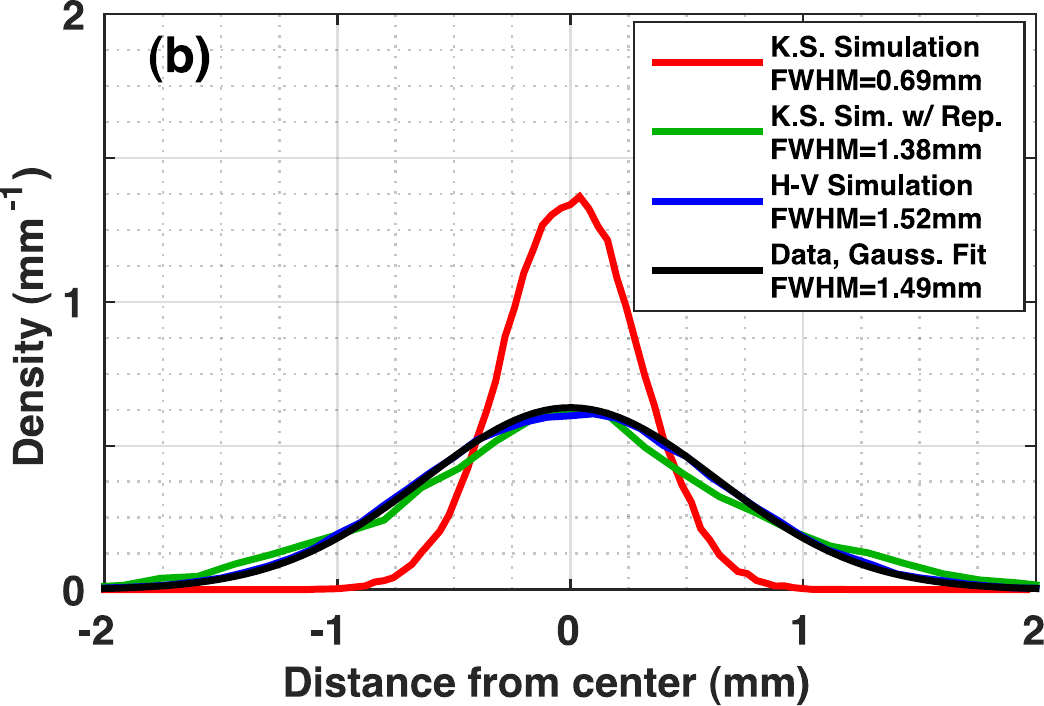}}
\caption{{\bf Comparison of Lateral Spreading of Electron Charge Density:} (a) and (b) show plots of one-dimensional charge density of a single electron cluster, projected along the narrow and wide principle axes, respectively. A Gaussian fit to the data is shown in black, and is compared to three different simulations. For both simulations and data, the electric field strength was 2V/cm.}
\label{fig:ElectronLateralDiffusion}
\vspace{-0.5cm}
\end{center}
\end{figure}


Figure \ref{fig:Inter-valleyScattering} shows the mean path length between electron inter-valley scattering events, projected along the $\left[ 0, 0, 1 \right]$ direction, as a function of electric field strength. The data from our experiment are shown in blue, and are in good agreement with the Redl simulation, shown in green. Note that the parameters which determine the electron scattering rate in the Redl simulation were taken from reference \cite{Leman}, and were not modified to fit our data. For comparison, we also show data from references \cite{Broniatowski2014} and \cite{Piro2014}, plotted on the same axes.


\section{Monte Carlo Simulation}

To obtain good agreement between the Monte Carlo simulation by Peter Redl and the experimental data, several effects need to be taken into account. The first effect is a distortion in the locations of the centers of the four electron clusters. This distortion is explained by a 2 degree offset between the crystal axes and the faces of the crystal, which is consistent with the tolerances of our crystal fabrication process and with subsequent X-ray diffraction measurements of the crystal axes.

The second effect is the lateral spreading of the carriers due to electrostatic repulsion. Without accounting for this effect, our original Monte Carlo simulation for the holes produced charge distributions that were too narrow (see Fig. \ref{fig:Holes}c). The additional spreading seen in our data can be reproduced in our simulations by incorporating the effect of electrostatic repulsion. Based on the total charge collected by the large electrode on the non-illuminated face, we estimate $1.9 \pm .4 \times 10^5$ charge carriers are collected per laser pulse.\cite{GeneratedCarriers} To obtain good agreement with the data, the simulation contains $2 \pm 1 \times 10^5$ holes, assuming the initial charge cloud diameter after the separation of opposite charges is equal in size to the laser spot (60$\mu$m).


However, even after accounting for electrostatic repulsion, a discrepancy remains in the case of the electrons, as the electron charge distribution is narrower in the data than in the Redl simulation (compare Figs. \ref{fig:Electrons}a \& \ref{fig:Electrons}b). We attribute this discrepancy to the fact that the Redl simulation uses the Herring-Vogt approximation.\cite{Herring1956,Leman} This approximation provides a 20$\times$ faster means of simulating the phonon emission process at the expense of violating momentum conservation. This non-conservation of momentum is likely the cause of the extra lateral diffusion seen in the simulated charge density patterns.

To estimate the systematic error introduced by the Herring-Vogt approximation, we use a simulation written by Kyle Sundqvist\cite{Sundqvist2012,Phipps2012,Sundqvist2014} which explicitly conserves crystal momentum. Figure \ref{fig:SimulationComparison} shows a comparison of the Redl simulation (a), with the more accurate, but slower, Sundqvist simulation (b). The widths of the distributions of electrons in a single cluster are shown in Fig. \ref{fig:ElectronLateralDiffusion} for both simulations, and for the data.


Using the Sundqvist simulation as a standard of comparison, we are able to observe and model the extra lateral spreading of the electrons due to electrostatic repulsion. The repulsion causes the centers of the four electron clusters to spread further from each other, and causes the clusters to spread more in the low-mass directions. Figure \ref{fig:SimulationComparison}c shows an approximate correction to the Sundqvist simulation for the effects of electrostatic repulsion, generated by post-processing the electron trajectories. Figure \ref{fig:ElectronLateralDiffusion} shows how this effect brings the Sundqvist simulation into agreement with the experimental data, assuming the same number of charge-carriers per pulse ($2 \pm 1 \times 10^5$) as we assume for holes.

We also note that, unlike the Redl simulation, the Sundqvist simulation does not include the effects of impurity scattering. This fact is apparent in Fig. \ref{fig:Inter-valleyScattering}, where the IV scattering length for the Sundqvist simulation diverges for low electric field. However, a discrepancy still exists between the Redl and Sundqvist simulations even at high fields where scattering is dominated by phonon emission. This discrepancy remains unexplained.


\section{Conclusions}


\vspace{-0.4cm}

Using our novel cryogenic scanning apparatus, we have, for the first time, imaged the two-dimensional charge density distributions of both electrons and holes in high purity germanium at sub-Kelvin temperatures. These results provide a useful check on the accuracy of our Monte Carlo simulations and provide motivation for improvements. Plans for future work include investigation of the scattering and trapping rates of charge carriers in multiple germanium and silicon crystals, which will improve our understanding of the low-temperature condensed-matter physics relevant to the design and analysis of the germanium and silicon detectors used by the CDMS collaboration for dark matter searches.

\begin{acknowledgments}

This work was supported in part by the U.S. Department of Energy and by the National Science Foundation. The authors are also especially grateful to the staff of the Varian Machine Shop at Stanford University for their assistance in machining the parts used in this experiment, and to members of Mirrorcle Technologies Inc., who provided a great deal of time and assistance in implementing the MEMS system, and modifying it to allow operation at temperatures below 1K.

\end{acknowledgments}



\end{document}